\newcommand{\xBc}{\langle}
\newcommand{\xBe}{\rangle}

\newcommand{\xbF}{\Phi}

\newcommand{\xba}{\alpha}

\newcommand{\xbe}{\in}
\newcommand{\xbf}{\phi}

\newcommand{\xbo}{\omega}

\newcommand{\xCN}{\neg}
\newcommand{\xCO}{ }
\newcommand{\xCQ}{\emptyset}

\newcommand{\xCf}{\hspace{0.1em}}

\newcommand{\xcA}{\forall}

\newcommand{\xcE}{\exists}

\newcommand{\xcT}{\bot}
\newcommand{\xcU}{\bigwedge}

\newcommand{\xcc}{\subseteq}

\newcommand{\xcg}{\geq}

\newcommand{\xck}{\leq}

\newcommand{\xco}{\vee}
\newcommand{\xcp}{\rightarrow}

\newcommand{\xcr}{\leftrightarrow}

\newcommand{\xct}{\top}
\newcommand{\xcu}{\wedge}
\newcommand{\xcv}{\cup}

\newcommand{\xcz}{\Box}

\newcommand{\xDH}{\item }

\newcommand{\xdZ}{\mbox{\boldmath$Z$}}

\newcommand{\xdl}{{\cal L}}

\newcommand{\xdx}{{\cal X}}

\newcommand{\xEI}{\begin{itemize}}
\newcommand{\xEJ}{\end{itemize}}

\newcommand{\xEd}{\neq}

\newcommand{\xEh}{\begin{enumerate}}

\newcommand{\xEj}{\end{enumerate}}

\newcommand{\xEn}{\begin{description}}
\newcommand{\xEp}{\end{description}}

\newcommand{\Xl}{\ldots}

\newcommand{\bl}{\begin{lemma} \rm}
\newcommand{\el}{\end{lemma}}
\newcommand{\br}{\begin{remark} \rm}
\newcommand{\er}{\end{remark}}
\newcommand{\be}{\begin{example} \rm}
\newcommand{\ee}{\end{example}}
\newcommand{\bco}{\begin{corollary} \rm}
\newcommand{\eco}{\end{corollary}}
\newcommand{\bc}{\begin{claim} \rm}
\newcommand{\ec}{\end{claim}}
\newcommand{\bfa}{\begin{fact} \rm}
\newcommand{\efa}{\end{fact}}
\newcommand{\bp}{\begin{proposition} \rm}
\newcommand{\ep}{\end{proposition}}
\newcommand{\bd}{\begin{definition} \rm}
\newcommand{\ed}{\end{definition}}
\newcommand{\bcs}{\begin{construction} \rm}
\newcommand{\ecs}{\end{construction}}
\newcommand{\bcd}{\begin{condition} \rm}
\newcommand{\ecd}{\end{condition}}
\newcommand{\bt}{\begin{theorem} \rm}
\newcommand{\et}{\end{theorem}}
\newcommand{\bn}{\begin{notation} \rm}
\newcommand{\en}{\end{notation}}
\newcommand{\bfi}{\begin{bild} \rm}
\newcommand{\efi}{\end{bild}}
\newcommand{\bsta}{\begin{statement} \rm}
\newcommand{\esta}{\end{statement}}
\newcommand{\bcom}{\begin{comment} \rm}
\newcommand{\ecom}{\end{comment}}
\newcommand{\bdia}{\begin{diagram} \rm}
\newcommand{\edia}{\end{diagram}}

\newcommand{\bfc}{\begin{figure}[htb] \begin{center}}
\newcommand{\efc}{\end{center} \end{figure}}

\documentclass{article}

\usepackage{amssymb,latexsym,epic,eepic,rotating}

\title{Remarks on an Article by Rabern et al.
\thanks{File: Yab
}
}

\author{Karl Schlechta
\thanks{
schcsg@gmail.com - https://sites.google.com/site/schlechtakarl/ -
Koppeweg 24, D-97833 Frammersbach, Germany}
\thanks{
Retired, formerly: Aix-Marseille Universit\'{e}, CNRS, LIF UMR 7279, F-13000
Marseille, France
}
}

\begin{document}

\newtheorem{lemma}{Lemma}[section]
\newtheorem{theorem}[lemma]{Theorem}
\newtheorem{proposition}[lemma]{Proposition}
\newtheorem{corollary}[lemma]{Corollary}
\newtheorem{claim}[lemma]{Claim}
\newtheorem{fact}[lemma]{Fact}
\newtheorem{remark}[lemma]{Remark}
\newtheorem{definition}{Definition}[section]
\newtheorem{construction}{Construction}[section]
\newtheorem{condition}{Condition}[section]
\newtheorem{example}{Example}[section]
\newtheorem{notation}{Notation}[section]
\newtheorem{bild}{Figure}[section]
\newtheorem{comment}{Comment}[section]
\newtheorem{statement}{Statement}[section]
\newtheorem{diagram}{Diagram}[section]

\renewcommand{\labelenumi}
  {(\arabic{enumi})}
\renewcommand{\labelenumii}
  {(\arabic{enumi}.\arabic{enumii})}
\renewcommand{\labelenumiii}
  {(\arabic{enumi}.\arabic{enumii}.\arabic{enumiii})}
\renewcommand{\labelenumiv}
  {(\arabic{enumi}.\arabic{enumii}.\arabic{enumiii}.\arabic{enumiv})}

\maketitle

\setcounter{secnumdepth}{3}
\setcounter{tocdepth}{3}

\begin{abstract}

We show that conjecture 15 in \cite{RRM13} is wrong, comment on
theorem 24 in \cite{RRM13}, and conclude with some remarks on structures similar
to the Yablo construction.

\end{abstract}

\tableofcontents
\clearpage

%
%
%
\section{
Introduction
}

This paper is a footnote to  \cite{RRM13}.
 \cite{RRM13} is perhaps best
described as a graph theoretical analysis of
Yablo's construction, see  \cite{Yab82}.
We continue this work.

To make the present paper self-contained, we repeat the
definitions of  \cite{RRM13}. To keep it short, we
do not repeat ideas and motivations of  \cite{RRM13}.
Thus, the reader should probably be familiar with or have a copy of
 \cite{RRM13} ready.

All graphs etc. considered will be assumed to be cycle-free, unless said
otherwise.
\subsection{
Overview
}

 \xEh
 \xDH
Section \ref{Section Definitions} (page \pageref{Section Definitions})  contains
most of the definitions we use,
many are taken from  \cite{RRM13}.
 \xDH
In Section 
\ref{Section Conjecture} (page 
\pageref{Section Conjecture}), we show that
conjecture 15 in  \cite{RRM13} is wrong.
This conjecture says that a directed graph $G$ is dangerous iff
every homomorphic image of $G$ is dangerous.
(The definitions are given in
Definition 
\ref{Definition Basics} (page 
\pageref{Definition Basics}), (3) and (11).)

To show that the conjecture is wrong, we
modify the Yablo construction, see
Definition 
\ref{Definition Yablo-Structure} (page 
\pageref{Definition Yablo-Structure}), slightly in
Example \ref{Example YG'} (page \pageref{Example YG'}), illustrated in
Diagram \ref{Diagram YG'} (page \pageref{Diagram YG'}), show that it is
still dangerous in Fact \ref{Fact YG'} (page \pageref{Fact YG'}),
and collaps it
to a homomorphic image in
Example \ref{Example Homomorphismus} (page \pageref{Example Homomorphismus}).
This homomorphic image is not dangerous, as
shown in Fact \ref{Fact Integer} (page \pageref{Fact Integer}).
 \xDH
In Section 
\ref{Section 24} (page 
\pageref{Section 24}), we discuss implications of Theorem 24
in  \cite{RRM13} - see the
paragraph immediately after the proof of the theorem
in  \cite{RRM13}.
This theorem states
that an undirected graph $G$ has a dangerous orientation iff it
contains a cycle.
(See Definition 
\ref{Definition Basics} (page 
\pageref{Definition Basics})  (4) for orientation.)

We show that for any simply connected directed graph $G$ - i.e., in the
underlying
undirected graph $U(G),$ from any two vertices $X,Y,$ there is at most one
path
from $X$ to $Y,$
see Definition 
\ref{Definition Simply} (page 
\pageref{Definition Simply})  - and for
any denotation $d$ for $G,$ we find an acceptable valuation for $G$ and
$d.$

The proof consists of a mixed induction, successively assigning values
for the $X,$ and splitting up the graph into ever smaller independent
subgraphs. The independence of the subgraphs relies essentially on the
fact that $G$ (and thus also all subgraphs of $G)$ is simply
connected.
 \xDH
In Section 
\ref{Section Various} (page 
\pageref{Section Various}), we discuss various modifications and
generalizations of the Yablo structure.
Remark \ref{Remark Yablo} (page \pageref{Remark Yablo})
illustrates the argument in the Yablo structure,
Example \ref{Example NotN} (page \pageref{Example NotN})
considers trivial modifications of the Yablo structure.
In Fact 
\ref{Fact Finite-Branching} (page 
\pageref{Fact Finite-Branching})  we show that
infinite branching is necessary for a graph
being dangerous, and
Example 
\ref{Example Only-Negative} (page 
\pageref{Example Only-Negative})  shows why
infinitely many finitely branching points cannot
replace infinite branching - there is an infinite
``procrastination branch''.

Finally, we define a generalization of the Yablo structure
in Definition 
\ref{Definition Yablo-Like} (page 
\pageref{Definition Yablo-Like}), a transitive graph, with all
$d(X)$ of the form $ \xcU \{ \xCN X_{i}:i \xbe I\}.$

Our main result here is in
Fact \ref{Fact TransAndNot} (page \pageref{Fact TransAndNot}), where we
show that in Yablo-like structures,
the existence of an acceptable valuation is strongly
related to existence of successor nodes,
where $X' $ is a successor of $X$ in a directed graph $G,$ iff $X \xcp X'
$ in $G,$
or, written differently, $XX' \xbe E(G),$ the set of edges in $G.$

 \xEj
\subsection{
Some definitions
}

\label{Section Definitions}

Notation and definitions are taken mostly from  \cite{RRM13}.

\bd

$\hspace{0.01em}$


\label{Definition Basics}

 \xEh
 \xDH
Given a (directed or not) graph $G,$ $V(G)$ will denote its set of
vertices, $E(G)$ its set of edges. In a directed graph, $xy \xbe E(G)$
will denote
an arrow from $x$ to $y,$ which we also write $x \xcp y,$
if $G$ is not directed, just a line from $x$ to $y.$

We often use $x,y,$ or $X,Y,$ etc. for vertices.
 \xDH
A graph $G$ is called transitive iff $xy,yz \xbe E(G)$ implies $xz \xbe
E(G).$
 \xDH
Given two directed graphs $G$ and $H,$ a homomorphism from $G$ to $H$ is a
function
$f:V(G) \xcp V(H)$ such that, if $xy \xbe E(G),$ then $f(x)f(y) \xbe
E(H).$
 \xDH
Given a directed graph $G,$ the underlying undirected graph is defined as
follows: $V(U(G)):=V(G),$ $xy \xbe E(U(G))$ iff $xy \xbe E(G)$ or $yx \xbe
E(G)),$ i.e., we
forget the orientation of the edges. Conversely, $G$ is called an
orientation of $U(G).$
 \xDH
$S,$ etc. will denote the set of propositional variables of some
propositional language $ \xdl,$ $S^{+},$ etc. the set of its formulas.
$ \xct $ and $ \xcT $ will be part of the formulas.
 \xDH
Given $ \xdl,$ $v$ will be a valuation, defined on $S,$ and extended to
$S^{+}$
as usual - the values will be $\{0,1\},$ $\{ \xct, \xcT \},$ or so.
$[s]_{v},$ $[ \xba ]_{v}$ will denote the valuation of $s \xbe S,$ $ \xba
\xbe S^{+},$ etc.
When the context is clear, we might omit the index $v.$
 \xDH
$d$ etc. will be a denotation assignment, or simply denotation, a function
from $S$ to $S^{+}.$
 \xDH
A valuation $v$ is acceptable on $S$ relative to $d,$ iff for all $s \xbe
S$
$[s]_{v}=[d(s)]_{v},$ i.e. iff $[s \xcr d(s)]_{v}= \xct.$
(When $S$ and $d$ are fixed, we just say that $v$ is acceptable.)
 \xDH
A system $(S,d)$ is called paradoxical iff there is no $v$ acceptable for
$S,$ $d.$
 \xDH
Given $S,$ $d,$ we define $G_{S,d}$ as follows:
$V(G_{S,d}):=S,$ $ss' \xbe E(G_{S,d})$ iff $s' \xbe S$ occurs in $d(s).$
 \xDH
A directed graph $G$ is dangerous iff there is a paradoxical system
$(S,d),$
such that $G$ is isomorphic to $G_{S,d}.$
 \xEj

\ed

\bd

$\hspace{0.01em}$


\label{Definition Downward}

Let $G$ be a directed graph, $x,x' \xbe V(G).$
 \xEh
 \xDH
$x' $ is a successor of $x$ iff $xx' \xbe E(G).$

$succ(x):=\{x':$ $x' $ is a successor of $x\},$
 \xDH
Call $x' $ downward from $x$ iff there is a path from $x$ to $x',$
i.e. $x' $ is in the transitive closure of the succ operator.
 \xDH
Let $[x \xcp ]$ be the subgraph of $G$ generated by $\{x\} \xcv \{x':x' $
is downward from $x\},$
i.e. $V([x \xcp ]):=\{x\} \xcv \{x':x' $ is downward from $x\},$ and $x'
\xcp x'' \xbe E([x \xcp ])$ iff
$x',x'' \xbe V([x \xcp ]),$ and $x' x'' \xbe E(G).$
 \xEj

\ed

\bd

$\hspace{0.01em}$


\label{Definition Yablo-Structure}

For easier reference, we define the Yablo structure,
see e.g.  \cite{RRM13}.

Let $V(G):=(Y_{i}:i< \xbo \},$ $E(G):=\{Y_{i}Y_{j}:i,j< \xbo,$ $i<j\},$
and
$d(Y_{i}):= \xcU \{ \xCN Y_{j}:i<j\}.$

$(Y_{i} \xbe S$ for a suitable language.)

\ed

\bd

$\hspace{0.01em}$


\label{Definition AndNot}

 \xEh
 \xDH
Call a denotation $d$ $ \xcU \xCN $ or $ \xcU -$ iff all $d(X)$ have the
form $d(X)= \xcU \{ \xCN X_{i}:i \xbe I\}$ -
as in the Yablo structure.
 \xDH
The dual notation $ \xcU +$ expresses the analogous case with $+$ instead
of $ \xCN,$
i.e. $d(X)= \xcU \{X_{i}:i \xbe I\}.$
 \xDH
We will use $ \xCN $ and - for negation, and $+$ when we want to emphasize
that
a formula is not negated.
 \xEj

\ed

\br

$\hspace{0.01em}$


\label{Remark Empty}

Note that we interpret $ \xcU $ in the strict sense of $ \xcA,$ i.e., $
\xCN \xcU \{ \xCN X_{i}:i \xbe I\}$
means that there is at least one $X_{i}$ which is true. In particular,
if $d(X)= \xcU \{ \xCN X_{i}:i \xbe I\},$ and $[X]=[d(X)]= \xcT,$ then
$d(X)$ must contain a propositional variable, i.e. cannot be composed only
of $ \xcT $ and $ \xct,$ so there is some arrow $X \xcp X' $ in the
graph.

Thus, if in the corresponding graph $succ(X)= \xCQ,$ $d$ is
of the form $ \xcU -,$ $v$ is an acceptable valuation for $d,$ then
$[X]_{v}= \xct.$

Dually, for $ \xcU +,$ $ \xCN \xcU \{X_{i}:i \xbe I\}$
means that there is at least one $X_{i}$ which is false.

Thus, if in the corresponding graph $succ(X)= \xCQ,$ $d$ is
of the form $ \xcU +,$ $v$ is an acceptable valuation for $d,$ then
$[X]_{v}= \xcT.$
\section{
A comment on conjecture 15 in \cite{RRM13}
}

\label{Section Conjecture}

\er

We show in this section that conjecture 15 in
 \cite{RRM13} is wrong.

\bd

$\hspace{0.01em}$


\label{Definition Contiguous}

Call $ \xdx \xcc \xdZ $ (the integers) contiguous iff for all $x,y,z \xbe
\xdZ,$ if $x<y<z$ and
$x,z \xbe \xdx,$ then $y \xbe \xdx,$ too.

\ed

\bfa

$\hspace{0.01em}$


\label{Fact Integer}

Let $G$ be a directed graph, $V(G)= \xdx $ for some contiguous $ \xdx,$
and
$x_{i}x_{j} \xbe E(G)$ iff $x_{j}$ is the direct successor of $x_{i}.$

Then for any denotation $d$:

 \xEh
 \xDH
$d(x)$ may be (equivalent to) $x+1,$ $ \xCN (x+1),$ $ \xcT,$ or $ \xct.$

If $d(x)= \xcT $ or $ \xct,$ we abbreviate $d(x)=c,$ $c',$ etc. (c for
constant).
 \xEj
If $v$ is acceptable for $d,$ then:
 \xEh
 \xDH
If $d(x)=c,$ then $d(x-1)=c' $ (if $x-1$ exists in $ \xdx).$
 \xDH
if $d(x)=(x+1),$ then $[x]_{v}=[x+1]_{v}$

if $d(x)= \xCN (x+1),$ then $[x]_{v}= \xCN [x+1]_{v}$
 \xDH
Thus:
 \xEh
 \xDH
If $d(x)=c$ for some $x,$ then for all $x' <x$ $d(x')=c' $ for some $c'
.$
 \xDH
We have three possible cases:
 \xEh
 \xDH
$d(x)=c$ for all $x \xbe \xdx,$
 \xDH
$d(x)=c$ for no $x \xbe \xdx,$
 \xDH
there is some maximal $x' $ s.t. $d(x')=c,$ so $d(x'') \xEd c' $ for all
$x'' >x'.$
 \xEj
 \xEI
 \xDH
In the first case, for all $x,$ if $d(x)$ is $ \xcT $ or $ \xct,$ then
the valuation
for $x$ starts anew, i.e. independent of $x+1,$ and continues to $x-1$
etc.
according to (2).
 \xDH
in the second case, there is just one acceptable valuation:
we chose some $x \xbe \xdx,$ and $[x]_{v}$ and propagate the value up and
down
according to (2)
 \xDH
in the third case, we work as in the first case up to $x',$ and treat the
$x'' >x' $ as in the second case.
 \xDH
Basically, we work downwards from constants, and up and down beyond the
maximal constant. Constants interrupt the upward movement.
 \xEJ
 \xEj
 \xDH
Consequently, any $d$ on $ \xdx $ has an acceptable valuation $v_{d},$ and
the graph
is not dangerous.

(The present fact is a special case of
Fact 
\ref{Fact SimplyConnected} (page 
\pageref{Fact SimplyConnected}), but it seems useful to discuss a
simple
case first.)
 \xEj

\efa

\be

$\hspace{0.01em}$


\label{Example YG'}

We define now a modified Yablo graph $ \xCf YG',$ and a corresponding
denotation $d,$
which is paradoxical.

We refer to Fig.3 in  \cite{RRM13}, and
Diagram \ref{Diagram YG'} (page \pageref{Diagram YG'}).

 \xEh
 \xDH
The vertices (and the set $S$ of language symbols):

We keep all $Y_{i}$ of Fig.3 in  \cite{RRM13},
and introduce new vertices $(Y_{i},Y_{j},Y_{k})$ for $i<k<j.$
(When we write $(Y_{i},Y_{j},Y_{k}),$ we tacitly assume that $i<k<j.)$
 \xDH
The arrows:

All $Y_{i} \xcp Y_{i+1}$ as before.
We ``factorize'' longer arrows through new vertices:
 \xEh
 \xDH
$Y_{i} \xcp (Y_{i},Y_{j},Y_{i+1})$
 \xDH
$(Y_{i},Y_{j},Y_{k}) \xcp (Y_{i},Y_{j},Y_{k+1})$
 \xDH
$(Y_{i},Y_{j},Y_{j-1}) \xcp Y_{j}$
 \xEj
See Diagram \ref{Diagram YG'} (page \pageref{Diagram YG'}).
 \xEj

We define $d$ (instead of writing $d((x,y,z))$ we write $d(x,y,z)$ -
likewise $[x,y,z]_{v}$ for $[(x,y,z)]_{v}$ below):
 \xEh
 \xDH
$d(Y_{i}):= \xCN Y_{i+1} \xcu \xcU \{ \xCN (Y_{i},Y_{j},Y_{i+1}):$ $i+2
\xck j\}$

(This is the main idea of the Yablo construction.)
 \xDH
$d(Y_{i},Y_{j},Y_{k})$ $:=$ $(Y_{i},Y_{j},Y_{k+1})$ for $i<k<j-1$
 \xDH
$d(Y_{i},Y_{j},Y_{j-1})$ $:=$ $Y_{j}$
 \xEj

Obviously, $ \xCf YG' $ corresponds to $S$ and $d,$ i.e. $YG' =G_{S,d}.$

\ee

\bfa

$\hspace{0.01em}$


\label{Fact YG'}

$ \xCf YG' $ and $d$ code the Yablo Paradox:

\efa

\subparagraph{
Proof
}

$\hspace{0.01em}$


Let $v$ be an acceptable valuation relative to $d.$

Suppose $[Y_{1}]_{v}= \xct,$ then $[Y_{2}]_{v}= \xcT,$ and
$[Y_{1},Y_{k},Y_{2}]_{v}= \xcT $ for $2<k,$ so
$[Y_{k}]_{v}= \xcT $ for $2<k,$ as in Fact 
\ref{Fact Integer} (page 
\pageref{Fact Integer}), (2).
By $[Y_{2}]_{v}= \xcT,$ there must be $j$ such that
$j=3$ and $[Y_{3}]_{v}= \xct,$ or $j>3$ and $[Y_{2},Y_{j},Y_{3}]_{v}=
\xct,$ and as in
Fact \ref{Fact Integer} (page \pageref{Fact Integer}), (2)
again, $[Y_{j}]_{v}= \xct,$
a contradiction.

If $[Y_{1}]_{v}= \xcT,$ then as above for $[Y_{2}]_{v},$ we find $j \xcg
2$ and $[Y_{j}]_{v}= \xct,$ and
argue with $Y_{j}$ as above for $Y_{1}.$

Thus, $ \xCf YG' $ with $d$ as above is paradoxical, and $ \xCf YG' $ is
dangerous.

$ \xcz $
\\[3ex]

\be

$\hspace{0.01em}$


\label{Example Homomorphismus}

We first define $ \xCf YG'' $:
$V(YG''):=\{ \xBc Y_{i} \xBe:i< \xbo \},$ $E(YG''):=\{ \xBc Y_{i} \xBe  \xcp 
\xBc Y_{i+1} \xBe:i<
\xbo \}.$

We now define the homomorphism from $ \xCf YG' $ to $ \xCf YG''.$
We collaps for fixed $k$ $Y_{k}$ and all $(Y_{i},Y_{j},Y_{k})$ to
$ \xBc Y_{k} \xBe,$ more precisely,
define $f$
by $f(Y_{k}):=f(Y_{i},Y_{j},Y_{k}):= \xBc Y_{k} \xBe $ for all suitable $i,j.$

Note that $ \xCf YG' $ only had arrows between ``successor levels'',
and we have now only arrows from $ \xBc Y_{k} \xBe $ to $ \xBc Y_{k+1} \xBe,$
so $f$ is a
homomorphism,
moreover, our structure $ \xCf YG'' $ has the
form described in Fact \ref{Fact Integer} (page \pageref{Fact Integer}),
and is not dangerous, contradicting
conjecture 15 in  \cite{RRM13}.

\clearpage

\begin{diagram}

\label{Diagram YG'}
\index{Diagram YG'}

\unitlength1.0mm
\begin{picture}(150,180)(0,0)

\put(0,175){{\rm\bf Diagram YG' }}

\put(0,160){$Y_1$}
\put(140,160){$ \xBc Y_1 \xBe $}
\put(2,158){\line(0,-1){32}}
\put(143,158){\line(0,-1){33}}
\put(3,158){\line(3,-4){25}}
\put(4,158){\line(4,-3){45}}
\put(5,158){\line(5,-2){80}}

\put(0,120){$Y_2$}
\put(2,118){\line(0,-1){32}}
\put(143,118){\line(0,-1){33}}
\put(20,120){$(Y_1,Y_3,Y_2)$}
\put(50,120){$(Y_1,Y_4,Y_2)$}
\put(80,120){$(Y_1,Y_5,Y_2)$}
\put(140,120){$ \xBc Y_2 \xBe $}
\put(3,118){\line(5,-2){80}}
\put(4,118){\line(3,-1){105}}
\put(27,118){\line(-3,-4){25}}
\put(55,118){\line(-3,-4){25}}
\put(85,118){\line(-3,-4){25}}

\put(0,80){$Y_3$}
\put(2,78){\line(0,-1){32}}
\put(143,78){\line(0,-1){33}}
\put(20,80){$(Y_1,Y_4,Y_3)$}
\put(50,80){$(Y_1,Y_5,Y_3)$}
\put(80,80){$(Y_2,Y_4,Y_3)$}
\put(110,80){$(Y_2,Y_5,Y_3)$}
\put(140,80){$ \xBc Y_3 \xBe $}
\put(3,78){\line(5,-2){80}}
\put(27,78){\line(-3,-4){25}}
\put(55,78){\line(-3,-4){25}}
\put(80,79){\line(-2,-1){73}}
\put(115,78){\line(-3,-2){50}}

\put(0,40){$Y_4$}
\put(2,38){\line(0,-1){32}}
\put(143,38){\line(0,-1){33}}
\put(20,40){$(Y_1,Y_5,Y_4)$}
\put(50,40){$(Y_2,Y_5,Y_4)$}
\put(80,40){$(Y_3,Y_5,Y_4)$}
\put(140,40){$ \xBc Y_4 \xBe $}
\put(27,38){\line(-3,-4){25}}
\put(56,38){\line(-3,-2){51}}
\put(80,39){\line(-2,-1){73}}

\put(0,0){$Y_5$}
\put(140,0){$ \xBc Y_5 \xBe $}



\end{picture}

\end{diagram}

\vspace{4mm}

\ee

This is just the start of the graph, it continues downward through
$ \xbo $ many levels.

The lines stand for downward pointing arrows.
The lines originating from the $Y_{i}$ correspond to the negative
lines in the original Yablo graph, all others are simple positive
lines, of the type $d(X)=X'.$

The left part of the drawing represents the graph YG', the
right hand part the collapsed graph, the homomorphic image YG".

Compare to Fig.3 in  \cite{RRM13}.
\clearpage
\section{
A comment on Theorem 24 of \cite{RRM13}
}

\label{Section 24}

We comment in this section on the meaning of theorem 24
in  \cite{RRM13}.

\bd

$\hspace{0.01em}$


\label{Definition Symbols}

Fix a denotation $d.$

Let $s(X):=s(d(X))$ be the set of $s \xbe S$ which occur in $d(X).$

Let $r(X) \xcc s(X)$ be the set of relevant $s,$ i.e. which influence
$[d(X)]_{v}$ for
some $v.$
E.g., in $(\xba \xco \xCN \xba) \xcu \xba',$ $ \xba' $ is relevant, $
\xba $ is not.

\ed

\bd

$\hspace{0.01em}$


\label{Definition Simply}

 \xEh
 \xDH
Let $G$ be a directed graph.
For $X \xbe G,$ let the subgraph $C(X)$ of $G$ be the connected component
of $G$ which
contains $X:$
$X \xbe V(C(X)),$ and
$X' \xbe V(C(X))$ iff there is a path in $U(G)$ from $X$ to $X',$
together with the induced edges of $G,$ i.e., if $Y,Y' \xbe V(C(X)),$ and
$YY' \xbe E(G),$
then $YY' \xbe E(C(X)).$
 \xDH
$G$ is called a simply connected graph iff for all $X,Y$ in $G,$ there is
at most
one path in $U(G)$ from $X$ to $Y.$

(One may debate if a loop $X \xcp X$ violates simple connectedness, as we
have the paths $X \xcp X$ and $X \xcp X \xcp X$ - we think so. Otherwise,
we exclude loops.)
 \xDH
Two subgraphs $G',$ $G'' $ of $G$ are disconnected iff there is no path
from any
$X' \xbe G' $ to any $X'' \xbe G'' $ in $U(G).$
 \xEj

\ed

\bfa

$\hspace{0.01em}$


\label{Fact Unconnected}

Let $G,$ $d$ be given, $G=G_{S,d}.$

If $G',$ $G'' $ are two disconnected subgraphs of $G,$ then they can be
given
truth values independently.

\efa

\subparagraph{
Proof
}

$\hspace{0.01em}$


Trivial, as the subgraphs share no propositional variables.
$ \xcz $
\\[3ex]

\bfa

$\hspace{0.01em}$


\label{Fact SimplyConnected}

Let $G$ be simply connected, and $d$ any denotation, $G=G_{S,d}.$
Then $G,d$ has an acceptable valuation.

\efa

\subparagraph{
Proof
}

$\hspace{0.01em}$


This procedure assigns an acceptable valuation to $G$ and $d$ in several
steps.

More precisely, it is an inductive procedure, defining $v$ for more and
more elements, and cutting up the graph into diconnected subgraphs.
If necessary, we will use unions for the definition of $v,$ and the common
refinement for the subgraphs in the limit step.

The first step is a local step, it tries to simplify $d(X)$ by looking
locally
at it, propagating [X] to $X' $ with $X' \xcp X$ if possible, and erasing
arrows
from and to $X,$ if possible. Erasing arrows decomposes the graph into
disconnected subgraphs, as the graph is simply connected.

The second step initializes an arbitrary value $X$ (or, in step (4), uses
a value determined in step (2)), propagates the value to $X' $ for $X'
\xcp X,$
erases the arrow $X' \xcp X.$ Initialising $X$ will have repercussions on
the
$X'' $ for $X \xcp X'',$ so we chose a correct possibility for the $X'' $
(e.g.,
if $d(X)=X'' \xcu X''',$ setting $[X]= \xct,$ requires to set $[X''
]=[X''' ]= \xct,$ too),
and erase the arrows $X \xcp X''.$ As $G$ is simply connected, the only
connection between the different $C(X'')$ is via $X,$ but this was
respected
and erased, and they are now independent.

 \xEh
 \xDH
Local step
 \xEh
 \xDH
For all $X' \xbe s(X)-r(X):$

 \xEh
 \xDH
replace $X' $ in $d(X)$ by $ \xct $ (or, equivalently, $ \xcT),$
resulting in
logically equivalent $d' (X)$ $(s(X')$ might now be empty),

 \xDH
erase the arrow $X \xcp X'.$

Note that $C(X')$ will then be disconnected from $C(X),$ as $G$ is simply
connected.
 \xEj
 \xDH
Do recursively:

If $s(d(X))= \xCQ,$ then $d(X)$ is equivalent to $ \xct $ (or $ \xcT)$
(it might also be
$ \xct \xcu \xcT $ etc.),
so $[X]_{v}=[d(X)]_{v}= \xct $ (or $ \xcT)$ in any acceptable valuation,
and $[d(X)]_{v}$
is independent of $v.$

 \xEh
 \xDH
For $X' \xcp X,$ replace $X$ in $d(X')$ by $ \xct $ (or $ \xcT)$ $(s(X'
)$ might now be empty),
 \xDH
erase $X' \xcp X$ in $G.$

$X$ is then an isolated point in $G,$ so its truth value is independent
of the other truth values (and determined already).
 \xEj
 \xEj

 \xDH
Let $G'' $ be a non-trivial (i.e. not an isolated point) connected
component
of the original graph $G,$ chose $X$ in $G''.$
If $X$ were already fixed as $ \xct $ or $ \xcT,$ then $X$ would have
been isolated
by step (1). So $[X]_{v}$ is undetermined so far.
Moreover, if $X \xcp X' $ in $G'',$ then $d(X')$ cannot
be equivalent to a constant value either, otherwise, the arrow $X \xcp X'
$ would
have been eliminated already in step (1).

Chose arbitrarily a truth value for $d(X),$ say $ \xct.$

 \xEh
 \xDH
Consider any $X' $ s.t. $X' \xcp X$ (if this exists)
 \xEh
 \xDH
Replace $X$ in $d(X')$ with that truth value, here $ \xct.$

 \xDH
Erase $X' \xcp X$

As $G'' $ is simply connected, all such $C(X')$ and $C(X)$ are now
mutually
disconnected.
 \xEj
 \xDH
Consider simultanously all $X'' $ s.t. $X \xcp X''.$
(They are not constants, as any $X'' \xbe V(G'')$ must be a propositional
variable.)

 \xEh
 \xDH
Chose values for all such $X'',$ corresponding to $[X]_{v}=[d(X)]_{v}$
$(= \xct $ here).

E.g., if $d(X)=X'' \xcu X''',$ and the value for $X$ was $ \xct,$ then
we have to chose
$ \xct $ also for $X'' $ and $X'''.$

This is possible independently by
Fact \ref{Fact Unconnected} (page \pageref{Fact Unconnected}),
as the graph $G'' $ is simply connected, and
$X$ is the only connection between the different $X'' $
 \xDH
Erase all such $X \xcp X''.$

$X$ is now an isolated point, and
as $G'' $ is simply connected, all $C(X'')$ are mutually disconnected,
and disconnected
from all $C(X')$ with $X' \xcp X,$ considered in (2.1).
 \xEj
 \xEj
The main argument here is that we may define $[X'' ]_{v}$ and $[X'''
]_{v}$ for all
$X \xcp X'' $ and $X \xcp X''' $ independently, if we respect the
dependencies resulting
through $X.$
 \xDH
Repeat step (1) recursively on all mutually disconnected fragments
resulting
from step (2).
 \xDH
Repeat step (2) for all $X'' $ in (2.2), but instead of the free choice
for $[X]_{v}$ in (2), the choice for the $X'' $
has already been made in step (2.2.1),
and work with this choice.
 \xEj
$ \xcz $
\\[3ex]
\section{
Various remarks on the Yablo structure
}

\label{Section Various}

We comment in this section on various modifications and generalizations of
the Yablo structure. We think that the
transitivity of the graph, and the form of the $d(X)= \xcU \{ \xCN X_{i}:i
\xbe I\}$ are
the essential properties of ``Yablo-like'' structures.

We make this official:

\bd

$\hspace{0.01em}$


\label{Definition Yablo-Like}

A structure $G,$ $d$ is called Yablo-like iff $G$ is transitive, and
$d$ of the $ \xcU -$ form.

(See Remark \ref{Remark Empty} (page \pageref{Remark Empty})  for
our interpretation of $ \xcU.)$

\ed

\br

$\hspace{0.01em}$


\label{Remark Yablo}

This remark is for illustration and intuition.

In the Yablo structure,
after some $Y_{i}$ which is true, all $Y_{j},$ $j>i$ have to be false.
After some $Y_{i}$
which is false, there has to be some $Y_{j},$ $j>i$ which is true.

 \xEh
 \xDH
We can summarize this as the following two rules:
 \xEh
 \xDH
After $+,$ only - may follow, abbreviating:

after some $Y_{i}$ with $[Y_{i}]_{v}= \xct,$ all $Y_{j}$ with $j>i$ have
to be $[Y_{j}]_{v}= \xcT $
 \xDH
After -, there has to be some $+.$
 \xEj
 \xEI
 \xDH
This has only finite solutions: $- \Xl.-+,$ a sequence of -, ending with
last
element $+.$

Last-but-one $+$ does not work, as then the last is -, but we need an $+$
after
this one.
 \xDH
If we have an infinite sequence, there has to be a $+$ somewhere, followed
by
- only, contradiction.
 \xEh
 \xDH
if we start with $+,$ then the first - imposes a $+$ somewhere,
contradiction
 \xDH
if we start with -, then there has to be $+$ somewhere, say at element
$i,$
so $i+1$ has to be -, so some $j>i+1$ has to be $+,$ contradiction.
 \xEj
 \xEJ
 \xDH
An alternative view is the following:

$ \xcA $ (or $+)$ constructs defensive walls, $ \xcE $ (or -) attacks
them.

The elements of the walls (-) themselves are attacks on later parts of the
walls, the attacks attack earlier constructions $(+)$ of the walls.
 \xEj

\er

\be

$\hspace{0.01em}$


\label{Example NotN}

We discuss here some very simple examples, all modifications of the
Yablo structure.

Up to now, we considered graphs isomorphic to (parts of) the natural
numbers
with arrows pointing to bigger numbers. We consider now other cases.
 \xEh
 \xDH
Consider the negative numbers (with 0), arrows pointing again to
bigger numbers.
Putting $+$ at 0, and - to all other elements is an acceptable valuation.
 \xDH
Consider a tree with arrows pointing to the root. The tree may be
infinite.
Again $+$ at the root, - at all other elements is an acceptable valuation.
 \xDH
Consider an infinite tree, the root with $ \xbo $ successors $x_{i},$ $i<
\xbo,$ and
from each $x_{i}$ originating a chain of length $i$ as in Fig. 10
of  \cite{RRM13}, putting $+$ at the end of the branches, and -
everywhere else
is an acceptable valuation.
 \xDH
This trivial example shows that an initial segment of a Yablo construction
can again be a Yablo construction.

Instead of considering all $Y_{i},$ $i< \xbo,$ we consider $Y_{i},$ $i<
\xbo + \xbo,$
extending the original construction in the obvious way.
 \xEj

\ee

\bfa

$\hspace{0.01em}$


\label{Fact Finite-Branching}

Let $G$ be loop free and finitely forward branching, i.e. for any $s,$
there are
only finitely many $s' $ such that $s \xcp s' $ in $G.$ Then $G$ is not
dangerous.

(d may be arbitrary, not necessarily of the $ \xcU -$ form.)

\efa

\subparagraph{
Proof
}

$\hspace{0.01em}$


Let $d$ be any assignment corresponding to $G.$ Then $d(s)$ is a finite,
classical
formula. Replace $[s]_{v}=[d(s)]_{v}$ by the classical formula $
\xbf_{s}:=s \xcr d(s).$
Then any finite number of $ \xbf_{s}$ is consistent.

Proof: Let $ \xbF $ be a finite set of such $ \xbf_{s},$ and $S_{ \xbF }$
the set of $s$ occurring
in $ \xbF.$ As $G$ is loop free, and $S_{ \xbF }$ finite, we may
initialise the minimal $s \xbe S_{ \xbF }$ (i.e. there is no $s' $
such that $s \xcp s' $ in the part of $G$ corresponding to $ \xbF)$ with
any truth values, and propagate the truth
values upward according to usual valuation rules.
This shows that $ \xbF $ is consistent, i.e. we have constructed a
(partial) acceptable valuation for $d.$

Extend $ \xbF $ by classical compactness, resulting in a total acceptable
valuation for $d.$

(In general, in the logics considered here, compactness obviously does not
hold:
Consider $\{ \xCN \xcU \{Y_{i}:i< \xbo \} \xcv \{Y_{i}:i< \xbo \}.$
Clearly, every finite subset is
consistent, but the entire set is not.)

$ \xcz $
\\[3ex]

The following modification of the Yablo structure has only one acceptable
valuation for $Y_{1}:$

\be

$\hspace{0.01em}$


\label{Example Only-Negative}

Let $Y_{i},$ $i< \xbo $ as usual, and introduce new $X_{i},$ $3 \xck i<
\xbo.$

Let $Y_{i} \xcp Y_{i+1},$ $Y_{i} \xcp X_{i+2},$ $X_{i} \xcp Y_{i},$ $X_{i}
\xcp X_{i+1},$ with

$d(Y_{i}):=$ $ \xCN Y_{i+1} \xcu X_{i+2},$ $d(X_{i}):= \xCN Y_{i} \xcu
X_{i+1}.$

If $Y_{1}= \xct,$ then $ \xCN Y_{2} \xcu X_{3},$ by $X_{3},$ $ \xCN Y_{3}
\xcu X_{4},$ so, generally,

if $Y_{i}= \xct,$ then $\{ \xCN Y_{j}:$ $i<j\}$ and $\{X_{j}:$ $i+1<j\}.$

If $ \xCN Y_{1},$ then $Y_{2} \xco \xCN X_{3},$ so if $ \xCN X_{3},$
$Y_{3} \xco \xCN X_{4},$ etc., so, generally,

if $ \xCN Y_{i},$ then $ \xcE j(i<j,$ $Y_{j})$ or $ \xcA j\{ \xCN X_{j}:$
$i+1<j\}.$

Suppose now $Y_{1}= \xct,$ then $X_{j}$ for all $2<j,$ and $ \xCN Y_{j}$
for all $1<j.$
By $ \xCN Y_{2}$ there is $j,$ $2<j,$ and $Y_{j},$ a contradiction, or $
\xCN X_{j}$ for all $3<j,$
again a contradiction.

But $ \xCN Y_{1}$ is possible, by setting $ \xCN Y_{i}$ and $ \xCN X_{i}$
for all $i.$

Thus, replacing infinite branching by an infinite number of finite
branching does not work for the Yablo construction, as we can always chose
the
``procrastinating'' branch.

See Diagram \ref{Diagram Only-Negative} (page \pageref{Diagram Only-Negative}).

\clearpage

\begin{diagram}

\label{Diagram Only-Negative}
\index{Diagram Only-Negative}

\unitlength1.0mm
\begin{picture}(150,180)(0,0)

\put(0,175){{\rm\bf Diagram for Example \ref{Example Only-Negative}}}

\put(0,160){$Y_1$}
\put(2,158){\line(0,-1){32}}
\put(3,158){\line(3,-4){23}}
\put(0,132){\line(1,0){4}}

\put(0,120){$Y_2$}
\put(2,118){\line(0,-1){32}}
\put(25,120){$X_3$}
\put(27,118){\line(0,-1){32}}
\put(3,118){\line(3,-4){23}}
\put(26,118){\line(-3,-4){23}}
\put(0,92){\line(1,0){4}}
\put(6,94){\line(4,-3){3}}

\put(0,80){$Y_3$}
\put(2,78){\line(0,-1){32}}
\put(25,80){$X_4$}
\put(27,78){\line(0,-1){32}}
\put(3,78){\line(3,-4){23}}
\put(26,78){\line(-3,-4){23}}
\put(0,52){\line(1,0){4}}
\put(6,54){\line(4,-3){3}}

\put(0,40){$Y_4$}
\put(2,38){\line(0,-1){32}}
\put(25,40){$X_5$}
\put(27,38){\line(0,-1){32}}
\put(3,38){\line(3,-4){23}}
\put(26,38){\line(-3,-4){23}}
\put(0,12){\line(1,0){4}}
\put(6,14){\line(4,-3){3}}

\put(0,0){$Y_5$}
\put(25,0){$X_6$}

\end{picture}

\end{diagram}

\vspace{4mm}

\ee

The lines stand again for downward pointing arrows.
Crossed lines indicate negations.
\clearpage

\bfa

$\hspace{0.01em}$


\label{Fact TransAndNot}

Let $G$ be transitive, and $d$ be of the type $ \xcU -.$
 \xEh
 \xDH
If $ \xcE X.$ $(succ(X) \xEd \xCQ $ and $ \xcA X' \xbe succ(X).succ(X')
\xEd \xCQ),$ then
$d$ has no acceptable valuation.

Let acceptable $v$ be given, $[.]$ is for this $v.$

Case 1: $[X]= \xct.$ So for all $X' \xbe succ(X)$ $[X' ]= \xcT,$ and
there is such $X',$ so
(either by the prerequisite $succ(X') \xEd \xCQ,$ or by
Remark \ref{Remark Empty} (page \pageref{Remark Empty}))
$ \xcE $ $X'' \xbe succ(X').[X'' ]= \xct,$ but $succ(X') \xcc succ(X),$
a contradiction.

In abbreviation:
$X^{+}$ $ \xcp_{ \xcU -}$ $X'^{-}$ $ \xcp_{ \xcU -}$ $X''^{+}$

Case 2: $[X]= \xcT.$ So $ \xcE X' \xbe succ(X).[X' ]= \xct,$ so $ \xcA
X'' \xbe succ(X').[X'' ]= \xcT,$
and by prerequisite $succ(X') \xEd \xCQ,$ so there is such $X'',$
so by Remark 
\ref{Remark Empty} (page 
\pageref{Remark Empty})  $succ(X'') \xEd \xCQ,$ so
$ \xcE $ $X''' \xbe succ(X'').[X''' ]= \xct,$ but $succ(X'') \xcc
succ(X'),$ a contradiction.

$X^{-}$ $ \xcp_{ \xcU -}$ $X'^{+}$ $ \xcp_{ \xcU -}$ $X''^{-}$ $ \xcp_{
\xcU -}$ $X'''^{+}$

(Here we need Remark \ref{Remark Empty} (page \pageref{Remark Empty})
for the additional step from $X'' $ to $X''' $.)

 \xDH
Conversely:

Let
$ \xcA X$ $(succ(X)= \xCQ $ or $ \xcE X' \xbe succ(X).succ(X')= \xCQ):$

By Remark \ref{Remark Empty} (page \pageref{Remark Empty}),
if $succ(Y)= \xCQ,$ then for any
acceptable valuation, $[Y]= \xct.$ Thus, if there is $X' \xbe succ(X),$
$succ(X')= \xCQ,$
$[X' ]= \xct,$ and $[X]= \xcT.$

Thus, the valuation defined by $[X]= \xct $ iff $succ(X)= \xCQ,$ and $
\xcT $ otherwise is
an acceptable valuation. (Obviously, this definition is free from
contradictions.)
 \xEj

$ \xcz $
\\[3ex]

\efa

$ \xCO $

\end{document}